# A multidisciplinary approach to study precipitation kinetics and hardening in an Al-4Cu (wt. %) alloy

A. Rodriguez-Veiga[1], B. Bellón[1,2], I. Papadimitriou[1],

G. Esteban-Manzanares[1,2], I. Sabirov[1], J. Llorca[1,2]

[1] IMDEA Materials Institute, Calle Eric Kandel 2, Getafe, 28906, Madrid, Spain

[2] Department of Materials Science, Polytechnic University of Madrid/Universidad Politécnica de Madrid, E.T.S. de Ingenieros de Caminos, 28040, Madrid, Spain

*Abstract*

A multidisciplinary approach is presented to analyse the precipitation process in a model Al-Cu alloy. Although this topic has been extensively studied in the past, most of the investigations are focussed either on transmission electron microscopy or on thermal analysis of the processes. The information obtained from these techniques cannot, however, provide a coherent picture of all the complex transformations that take place during decomposition of supersaturated solid solution. Thermal analysis, high resolution dilatometry, (high resolution) transmission electron microscopy and density functional calculations are combined to study precipitation kinetics, interfacial energies, and the effect of second phase precipitates on the mechanical strength of the alloy. Data on both the coherent and semi-coherent orientations of the $\theta''/Al$ interface are reported for the first time. The combination of the different characterization and modelling techniques provides a detailed picture of the precipitation phenomena that take place during aging and of the different contributions to the strength of the alloy. This strategy can be used to analyse and design more complex alloys.

*Keywords: Al alloy, precipitation kinetics, high resolution transmission electron microscopy, thermal analysis, precipitation strengthening, ab initio calculations.*

# 1. Introduction

Precipitation hardening is the main strategy to increase the mechanical strength of the age-hardenable Al alloys [1]. Size, volume fraction and morphology of second phase precipitates determine the amount of precipitation strengthening. They are controlled by the parameters of the aging treatment, namely temperature and time, which govern the precipitation kinetics of each alloy. The precipitation process is a very complex phenomenon involving the decomposition of a supersaturated solid solution and the formation of a series of metastable precipitates at different stages of aging [1]. Numerous publications focused on precipitation kinetics in the age-hardenable Al alloys can be found in the literature and binary Al-Cu alloys stand among the most widely studied. The vast majority of these works focus entirely on either (high resolution) transmission electron microscopy (TEM) studies of the second phase precipitates formed during aging as a function of temperature and time [2-7], or thermal analysis of the precipitation process using differential scanning calorimetry (DSC) and/or dilatometry [8-12]. Only in limited number of papers, TEM studies were combined with simulation to describe/predict the precipitation process [4, 13]. Each characterization method has its own advantages, though it is very far from being sufficient to understand the complex precipitation process.

The main objective of the present work is to tackle precipitation kinetics in a classical Al-4Cu (wt. %) alloy using a multidisciplinary approach. Advanced techniques for thermal analysis, microstructural characterization and *ab initio* simulation are combined to study precipitation kinetics and the effect of second phase precipitates on mechanical strength. The information obtained from different characterization and simulation techniques provides a coherent picture of the sequence of the different phenomena that take place during aging and of the different contributions to the strength of the alloy. This strategy can be used to analyse and design more complex alloys.

# 2. Material and experimental procedures

## 2.1. Processing and heat treatments

The Al-4Cu (wt. %) alloy was selected for this investigation. The material was prepared by casting in an induction furnace (VSG 002 DS, PVA TePla) using high-purity Al granulates (99.9 wt. %) and fine Cu powders (purity 99.9 wt.%). The base components were melted in an alumina crucible in Ar atmosphere. The molten alloy was homogenized for 15 min and poured into a stainless steel mould with 4 cylindrical cavities of 12 mm in diameter and 130 mm in length. After casting, the cylindrical bars were homogenized for 24 h at 813K and subjected to a step solution heat treatment (6 hours at 723K followed by 6 hours at 773K and 10 hours at 813K) in a muffle furnace Carbolite CWF1300 to reduce the segregation effects derived from the solidification process. Afterwards, the bars were quenched in cold water at 293K to supress any possible phase transformation during cooling. The homogenized samples were subjected to natural aging for 504 h at ambient temperature (294K) and artificial aging at 453K for various periods of time (6, 18, 30, 72, 120 and 168 h) to study the influence of aging time on the microstructure and the microhardness. After each treatment, specimens were quenched in water at 293K to avoid additional precipitation.

## 2.2. Microhardness testing

Samples of about 2 cm × 1 cm × 0.5 cm were machined from the central region of the as-cast ingots. One surface was carefully ground and polished to mirror-like conditions using standard metallographic techniques. Microhardness of the specimens was evaluated using a Shimadzu HMV-2T Vickers tester under a load of 4.98 N and dwell time of 10 s. Microhardness measurements were reported as average values of 10 different tests.

## 2.3. Microstructural characterization

Specimens for TEM studies were ground to a thickness of about 100 μm and punched to produce circular discs of 3 mm in diameter. These foils were jet electropolished by using a chemical solution of 30% nitric acid in methanol (vol. %) at ≈ 243K. The precipitate structure of the alloy was characterized using a FEI Talos TEM operating at 200 KV. The different types of precipitates were disk-shaped and oriented parallel to the {100} planes of the FCC in the α-Al matrix, leading to three different orientation variants. High-resolution images were accomplished by using the $\langle 100 \rangle_\alpha$ centred high-angle annular dark-field mode (HAADF-STEM). In this orientation, the habit planes of two variants of the precipitates are roughly parallel to the electron beam, as shown in Fig. 1 [3]. Reciprocal lattice images were obtained using high resolution dark field TEM images along with fast Fourier transform (FFT) of the individual precipitates. The thickness of the foils in the beam direction was determined by measuring the spacing of Kossel- Möllnstedt fringes in the $\{022\}_\alpha$ in a $\langle 100 \rangle_\alpha$ two-beam convergent beam electron diffraction pattern [14].

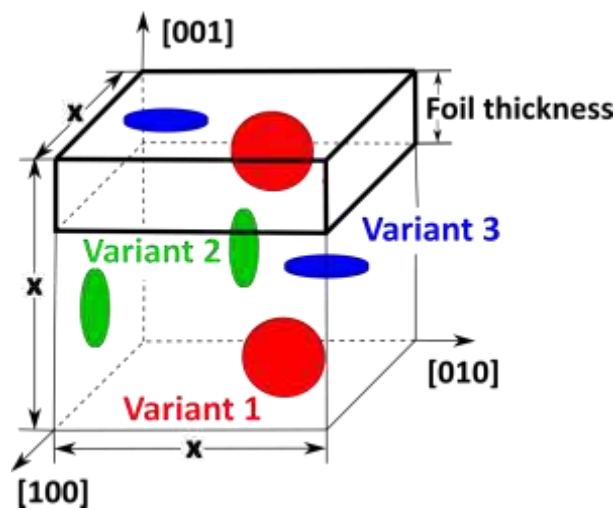

Figure 1 – Schematic representation of the precipitate distribution in a cubic volume of the matrix [3].

The apparent diameter $d_a$ and the thickness $t$ were measured in 50-100 particles in the thin TEM foils of thickness $\delta$ following the procedure proposed by Nie and Muddle [3]. Assuming that the precipitates have circular shape, the true precipitate diameter $d$ can be determined as,

$$d_a = \left(\frac{0.25\pi d + \delta}{\delta + d}\right) d \tag{1}$$

and the number density of precipitates is given by

$$n_p = \frac{N_T}{A_s(d+\delta)} = \frac{N_1 + N_2 + N_3}{A_s(d+\delta)} \tag{2}$$

where $N_1$, $N_2$ and $N_3$ designate the number of precipitates for the three different orientation variants, respectively, and $A_s$ stands for the area of the selected image. Orientation variants 1 and 2 show habit planes parallel to the electron beam (Fig. 1) and are clearly distinguishable in the STEM images. Nevertheless, the habit plane of the third type of variant of precipitates is perpendicular to the direction of the beam and they cannot be distinguished in the STEM micrographs. Following Nie and Muddle [3], $N_3$ was estimated as

$$N_3 = \left(\frac{N_1 + N_2}{2}\right)\left(\frac{d+\delta}{\sqrt{A_s}}\right) \tag{3}$$

Finally, the volume fraction of precipitates is given by

$$f = n_p \frac{\pi d^2 t}{4} \tag{4}$$

**2.4. Thermal analysis**

Two different techniques were used to study the aging kinetics of the alloy: differential scanning calorimetry (for continuous heating studies) and dilatometry (for isothermal studies).

*2.4.1. Differential scanning calorimetry analysis*

Specimens for thermal analysis were machined from the as-cast cylindrical ingots. Cubic samples having mass between 10-20 mg were fabricated for DSC analysis. The DSC experiments were performed using a TA-Q200 system and a $N_2$ protective atmosphere. DSC thermograms were recorded for specimens which have been solution treated, naturally aged and artificially aged for 6, 30 and 168 h. Samples were heated up to 773K at four different heating rates (2.5K/min, 5K/min, 10K/min and 20K/min, respectively). The DSC thermograms were analysed by means of the Kissinger peak reaction rate method [8] to determine the activation energy of the processes involved in the precipitation sequence, $Q$, which is given by [8]:

$$\ln\left(\frac{T_p^2}{\beta}\right) = -\frac{Q}{RT_p} + C_1 \tag{5}$$

where $T_p$ is the temperature at which the transformation rate of the reaction is maximum (i.e. peak temperature), $R$ the gas constant, $\beta$ the heating rate of the calorimetry test, and $C_1$ a constant. The activation energy of the process can be obtained from the slope of the plot of $\ln(T_p^2/\beta)$ *vs.* $1/RT_p$.

*2.4.2. Dilatometry analysis*

Cylindrical specimens of 10 mm in length and 8 mm in diameter with parallel smooth faces were machined for dilatometry studies. They were solution heat treated and dilatometry experiments were conducted using a NETZSCH DIL402 Supreme dilatometer in an Ar protective atmosphere at ambient temperature and 453 K to analyse the precipitation process at both temperatures in terms of sample dilatation [9]. The resolution of dilatation measurements was ~1 nm.

## 3. Microstructural development during aging

### 3.1. Artificial aging at 453 K

The age hardening response of the alloy during the artificial aging at 453K is shown in Fig. 2. The evolution of the hardness with time shows three different stages. The microhardness increased rapidly with aging time in the under-aged condition until the maximum microhardness (about 110 HV) was attained at the peak-aged condition after 30 h. Afterwards, the microhardness slightly decreased with aging time up to 120 h and more abruptly for longer aging times, leading to the overaged condition. This is the typical response of the Al-Cu alloys subjected to artificial aging [15].

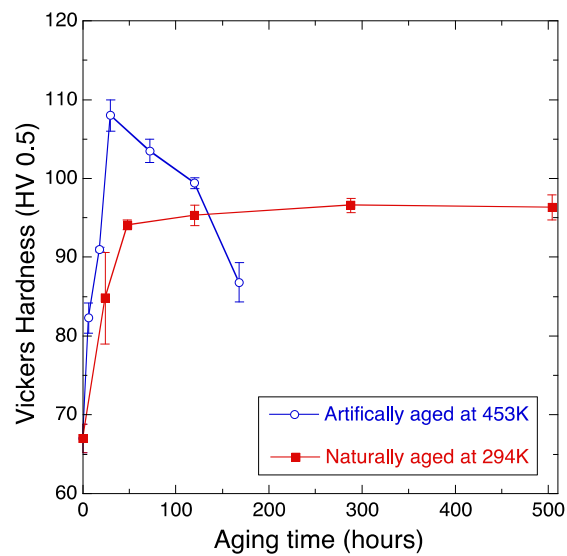

Figure 2 – Evolution of the microhardness of the Al-4%Cu alloy as a function of aging time at 294K (natural aging) and 453K (artificial aging). Error bars stand for the standard deviation corresponding to at least– 10 microhardness tests.

Three different kinds of precipitates, all of them with the shape of circular disks parallel to the {100} planes of the cubic FCC lattice of α-Al, were found by TEM in the samples aged at ambient and elevated temperature. Their structure is depicted in the high-resolution HAADF-STEM images in Fig. 3. Guinier-Preston (GP) zones, formed by a Cu-rich atom monolayer, are shown in Fig. 3(a). $\theta''$ precipitates were made up by 3 Al layers sandwiched between 2 Cu layers (Fig. 3b) and this sequence could be repeated twice (Fig. 3b) or, in rare occasions, more times (Fig. 3c), leading to multilayer $\theta''$ precipitates. Finally, thick $\theta'$ precipitates with a BCT structure formed after longer aging times at high temperature (Fig. 3d). Similar observations were reported in [4, 16-18].

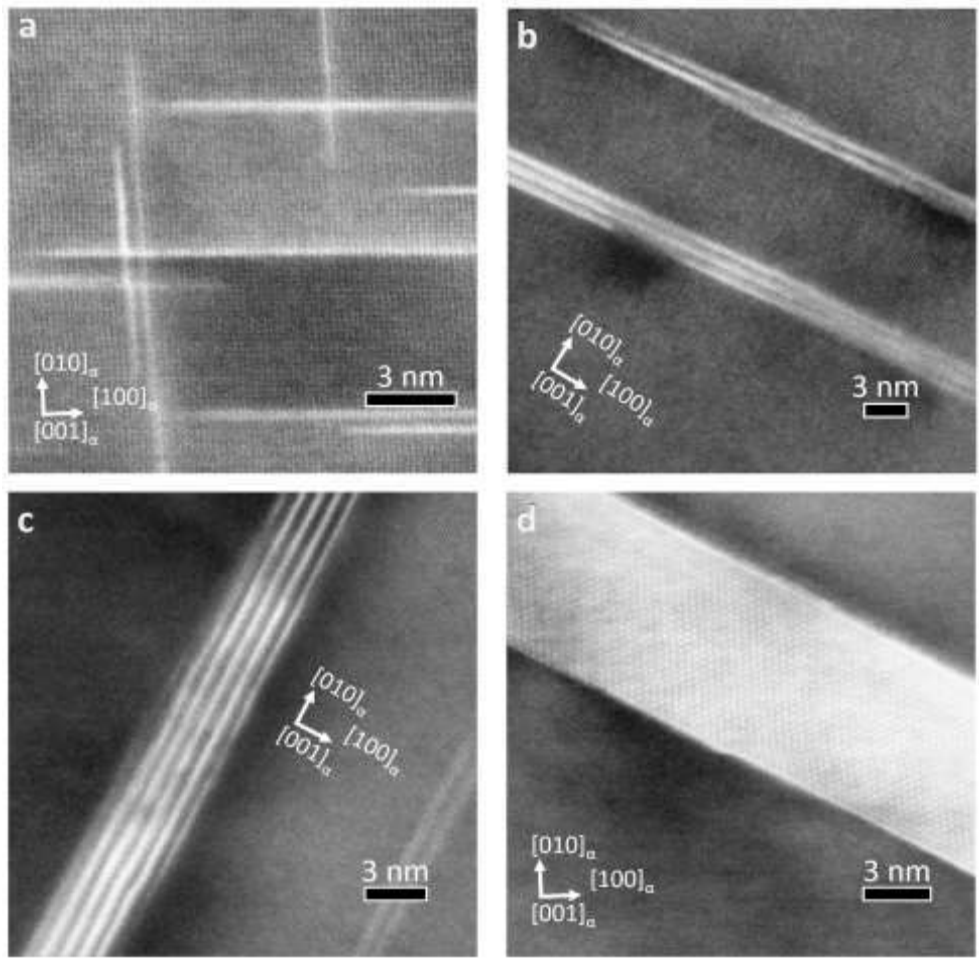

Figure 3 – High-resolution HAADF-STEM micrographs of the Al-4%Cu samples aged at 453K. (a) Monolayer GP zones formed after aging for 18 h. (b) $\theta''$ precipitates consisting of 3 Al layers sandwiched by 2 Cu layers formed after aging for 30 h. Sometimes this sequence was repeated twice. (c) Multilayer $\theta''$ precipitate formed after aging for 120 h and (d) thick $\theta'$ precipitate formed after aging for 168 h.

HAADF-STEM micrographs illustrating the spatial distribution of the precipitates during artificial aging are shown in Fig. 4. Images 4(a) and (b) reveal the precipitation structure of underaged samples, constituted by a distribution of fine GP zones and $\theta''$ precipitates [4, 16-17]. Several monolayer GP zones (as opposed to bilayer $\theta''$ precipitates) are marked with arrows in Fig. 4(a), which corresponds to 6 h of aging while several multilayers $\theta''$ precipitates are marked with arrows in Fig. 4(b), which corresponds to 18 h of aging.

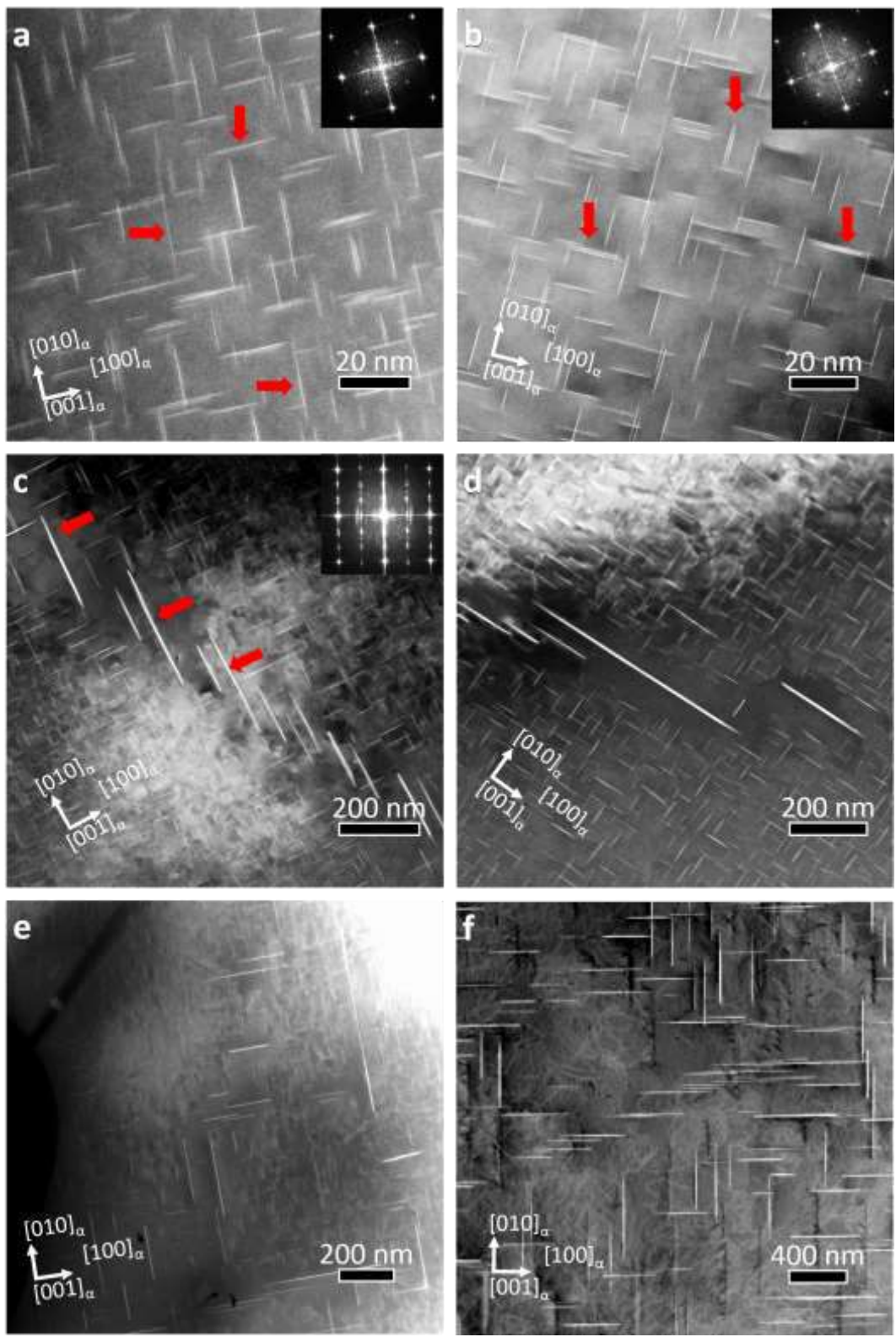

Figure 4 – HAADF-STEM micrographs of the Al-4%Cu specimens aged at 453 K for (a) 6 h. Several monolayer GP zones (as opposed to bilayer $\theta''$ precipitates) are marked with arrows. The FFT of the GP zones is shown in the inset. (b) 18 h. Several multilayers $\theta''$ precipitates are marked with arrows. The FFT of the $\theta''$ precipitates is shown in the inset. (c) 30 h. $\theta'$ precipitates nucleated on a dislocation forming a staircase structure are marked with arrows. The FFT of the $\theta'$ precipitates is shown in the inset. (d) 72 h. (e) 120 h and (f) 168 h. The electron beam was close to the $[001]_\alpha$ orientation.

Further aging up to 30 h leads to the peak-aged condition (Fig. 4c) whose microstructure is characterized by a uniform distribution of narrow $\theta''$ precipitates. $\theta'$ precipitates in the peak-aged condition were found to nucleate heterogeneously on grain boundaries and dislocations forming staircase-shaped structures as shown in Fig. 4(c). Further aging during 72 h (Fig. 4d) and 120 h (Fig. 4e) led to the homogeneous nucleation and growth of $\theta'$ precipitates, while $\theta'$ precipitates nucleated in dislocations grew and the volume fraction of $\theta''$ precipitates decreased. After longer aging times (168 h), the $\theta''$ phase was completely dissolved and the microstructure was dominated by a random and wide size distribution of thick $\theta'$ disks (Fig. 4f).

Table 1 - Quantitative measurements of the GP zones, $\theta''$ and $\theta'$ precipitates in the Al-4%Cu specimens naturally aged (NA) at 294K and artificially aged (AA) at 453K for different times. $\delta$ stands for the TEM foil thickness, $d$ the diameter, $t$ the thickness, $AR$ the aspect ratio, $n_p$ the number density and $f$ the volume fraction of the precipitates.

| Guinier Preston zones | | | | | | |
|---|---|---|---|---|---|---|
| Time (h) | $\delta$ (nm) | $d$ (nm) | $t$ (nm) | $AR$ | $n_p$ (µm$^{-3}$) | $f$ (%) |
| 504 (NA) | 124.5 | 2.5 ± 0.1 | 0.256 | 9.8 ± 0.5 | 7715913 ± 1313930 | 1.0 ± 0.3 |
| 18 (AA) | 133.9 | 14.6 ± 0.8 | 0.256 | 57.1 ± 3.3 | 92997 ± 9543 | 0.4 ± 0.1 |
| $\theta''$ precipitates | | | | | | |
| Time (h) | $\delta$ (nm) | $d$ (nm) | $t$ (nm) | $AR$ | $n_p$ (µm$^{-3}$) | $f$ (%) |
| 18 (AA) | 133.9 | 13.1 ± 0.6 | 1.4 ± 0.1 | 9.3 ± 0.8 | 13298 ± 1793 | 0.3 ± 0.1 |
| 30 (AA) | 123.1 | 31.3 ±1.4 | 1.6 ± 0.2 | 19.4 ± 2.8 | 15229 ± 1183 | 1.6 ± 0.4 |
| 120 (AA) | 127.5 | 45.1 ± 2.1 | 2.1 ± 0.2 | 21.6 ± 2.7 | 2883 ± 170 | 1.0 ± 0.2 |
| $\theta'$ precipitates | | | | | | |
| Time (h) | $\delta$ (nm) | d (nm) | $t$ (nm) | $AR$ | $n_p$ (µm$^{-3}$) | $f$ (%) |
| 30* (AA) | 123.1 | 171.9 ± 13.5 | 2.9 ± 0.2 | 59.5 ± 8.6 | 86 ± 11 | 0.6 ± 0.2 |
| 120 (AA) | 127.5 | 142.5 ± 25.1 | 4.9 ± 0.6 | 28.0 ± 2.4 | 131 ± 63 | 0.8 ± 0.1 |
| 168 (AA) | 173.4 | 342.5 ± 46.8 | 9.0 ± 0.6 | 39.0 ± 6.6 | 17 ± 6 | 1.0 ± 0.5 |

* Values measured in $\theta'$ plates that nucleated at dislocations

The average value and standard deviation of the precipitate diameter and thickness as well as of the number density and volume fraction are presented in Table 1 as a function of aging time. The thickness of the GP zones was always constant but the diameter after 18 h of aging at 453K was much longer than that measured after long time aging at ambient temperature. The $\theta''$ platelets appeared after 18 h and the average thickness (≈1.4 – 1.6 nm) was compatible with the sequence [CuAl$_3$]$_2$ according to the lattice parameters computed in the Appendix. As the aging time increased to 30 h, the diameter of the $\theta''$ platelets grew while the average thickness remained constant and thicker platelets compatible with the sequence [CuAl$_3$]$_3$ were only found after 120 h (Fig. 3c). $\theta'$ precipitates were only found after 30 h of aging and they were nucleated along dislocations, in agreement with recent multiscale simulations [13]. The large aspect ratio of these precipitates was a consequence of the interaction between the dislocation stress field and the stress-free transformation strain associated to the nucleation of the precipitate, which led to the formation of the staircase structures (Fig. 4c). Further aging up 120 h led to the homogeneous nucleation and growth of the $\theta'$ precipitates throughout the microstructure (Figs. 4e), with an aspect ratio of ≈ 20-30, in agreement with the multiscale simula-tions [13]. From 120 h to 168 h, the $\theta'$ precipitates grew in diameter and thickness while the aspect ratio remained constant.

The evolution of the volume fraction of each type of precipitates with aging time at 453K is plotted in Fig. 5. GP zones disappeared after 30 h of aging, while the number density and volume fraction of the $\theta''$ platelets increased up to the peak-aged condition at 30 h and decreased to zero after 168 h. The $\theta'$ precipitates only appeared after 30 h of aging. The volume fraction of $\theta'$ precipitates at 30 h was significant (≈ 0.6%), and they were mainly nucleated around dislocations and grain boundaries. Further aging promoted the homogeneous nucleation of $\theta'$ precipitates, whose volume fraction increased with aging time and reached 1.0% after 168 h. It should be noted, however, that the increase in the volume fraction of $\theta'$ precipitates between 120 h and 168 h was mainly due to the coarsening of the precipitates while the number of precipitates per unit volume decreased from ≈ 131 $\mu m^{-3}$ to 17 $\mu m^{-3}$ (Table 1).

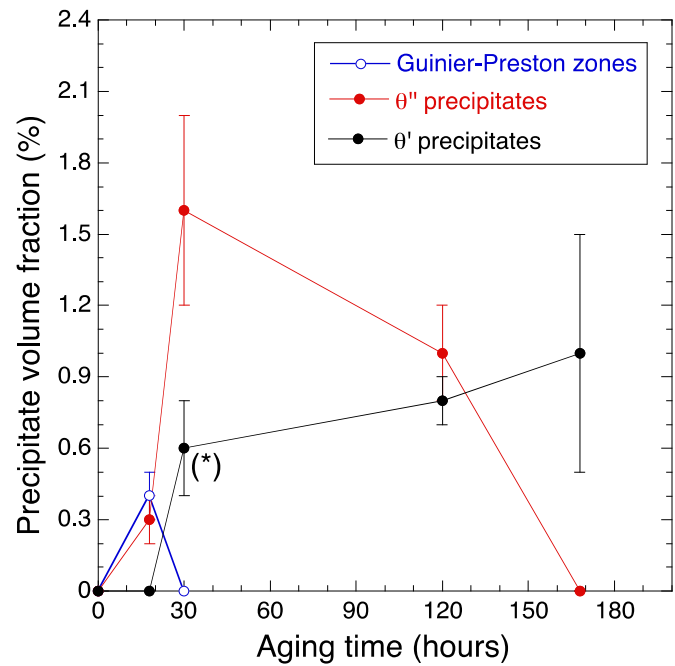

Figure 5 – Evolution of the volume fraction of precipitates with time during artificial aging at 453K. The (*) indicates that the $\theta'$ precipitates formed after 30 h at 453K were nucleated at dislocations and grain boundaries.

Taking into account TEM observations (Fig. 3–4), the microhardness curve (Fig. 2) and the quantitative measurements of the precipitate type, dimensions and volume fraction in Table 1 and Fig. 5, it can be suggested that the increase in hardness up to the peak-aged condition (30 h) has to be attributed to the homogeneous distribution of the $\theta''$ precipitates because the $\theta'$ precipitates in this stage were mainly clustered around dislocations and grain boundaries and did not contribute significantly to the strengthening of the alloy. The slight decrease of the microhardness at the second stage (between 30 and 120 h) could not be attributed to a reduction of the overall volume fraction of precipitates but to the substitution of small $\theta''$ precipitates by larger $\theta''$ (due to coarsening) and $\theta'$ precipitates. Finally, coarsening of $\theta'$ precipitates was responsible for the reduction in hardness after more 120 h of heat treatment in the overaged condition. Thus, it can be hypothesized that the bilayer and three-layer $\theta''$ disks are the most effective strengthening precipitates in the material (Table 1 and Fig. 2), and that increasing their volume fraction is the best strategy to enhance the hardening response of the alloy during artificial aging at 453K. In addition, increasing the number density or aspect ratio of particles does not necessarily improve the strength of the alloy. These hypotheses will be contrasted in section 5, in which the contribution of individual precipitates to the mechanical strength is analysed.

## 3.2. Natural aging

The microhardness increased with time in the naturally aged samples in the first 50 hours and reached an asymptotic value of around 95 HV after 300 h (Fig. 2). This microhardness is just 15 HV lower than the one measured at the peak-aged condition in the specimen aged at 453K and demonstrates the hardening potential of GP zones. The microstructure of the specimens naturally aged during 504 h is illustrated in the dark-field HAADF-STEM micrograph in Fig. 6. The precipitate structure is constituted by a uniform distribution of monolayer GP zones of ≈ 2.5 nm in diameter (Table 1). The thermodynamic driving force at room temperature is not sufficient to grow larger GP zones. No evidence of the precipitation of $\theta''$ platelets was found.

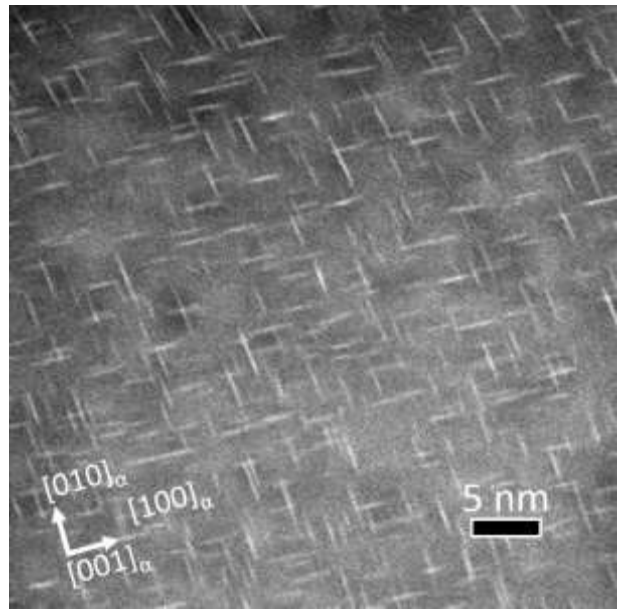

Figure 6 – HAADF-STEM micrograph of the Al-4%Cu specimen aged at ambient temperature for 504 h. The electron beam was close to the $[001]_\alpha$ orientation.

## 4. Precipitation kinetics

### 4.1. DSC analysis

The precipitation sequence in the binary Al-Cu alloys follows the sequence: supersaturated solid solution (SSSS) → Cu clusters → GP zones → $\theta''$ → $\theta'$ → $\theta$ [8, 12, 19]. Fig. 7 illustrates the calorimetry thermograms obtained by DSC in the Al-4%Cu samples after the solution heat treatment, natural aging at 294K for 504 h and artificial aging at 453K for different times. The thermograms in this figure were obtained at different heating rates, from 20K/min to 2.5K/min. The DSC scans were vertically translated for better comparison of the results, thus modifying the values of the ordinate axis corresponding to the heat flux. Nevertheless, both the position and the shape of the peaks were not modified. Eight peaks, four endothermic (A, C, E G) and four exothermic (B, D, F, H), were found in the thermograms at a heating rate of 20 K/min of the solution heat treated sample (Fig. 7a). Taking into account the precipitation sequence and earlier investigations [10, 11], the endothermic peaks were ascribed to the dissolution of the Cu-rich clusters (A), GP zones (C), $\theta''$ (E) and $\theta'$ precipitates (G). The exothermic peaks were associated with the precipitation of GP zones (B), $\theta''$ (D), $\theta'$ (F) and the equilibrium $\theta$ phase (H). Peaks associated with the formation of the quenched clusters and the dissolution of the $\theta$ precipitates were not detected in this study.

The thermograms in Fig. 7a clearly show that thermal history changes the precipitation kinetics. The first thermal effect observed in the thermogram of the solution treated alloy corresponds to the dissolution of the quenched Cu clusters (peak A) and this indicates that they were formed instantly after water quenching while the subsequent transformation of the quenched Cu-rich clusters into GP zones (peak B) occurred at the early stages of the aging treatment. These peaks were not found in the thermogram of the naturally aged specimen, in which the first peak observed (C) corresponded to the dissolution of the GP zones. These particles were formed during natural aging and reached a volume fraction of 1% with an average diameter of 2-3 nm (Table 1 and Fig. 6).

Artificial aging promoted the transformation of the GP zones into the $\theta''$ precipitates. However, neither the dissolution of the GP zones (peak C) nor the precipitation of the $\theta''$ phase (peak D) could be detected in the AA samples, not even at the highest heating rate (20K/min). It must be noted that the thermal trace of the processes defined by peaks A-D was already weak in the NA and SSSS conditions, probably due to the minuscule mass of the precipitates involved in such reactions. It is known that the size of DSC peaks depends on the energy released/absorbed by the specimen when an internal reaction takes place, i.e. the dissolution or precipitation of an intermetallic phase. The energy related to these processes increases with the heating rate [10]. Increasing the heating rates above 20 K/min led, however, to very large shifts of the peak position and the thermograms could not be used to identify the different reactions.

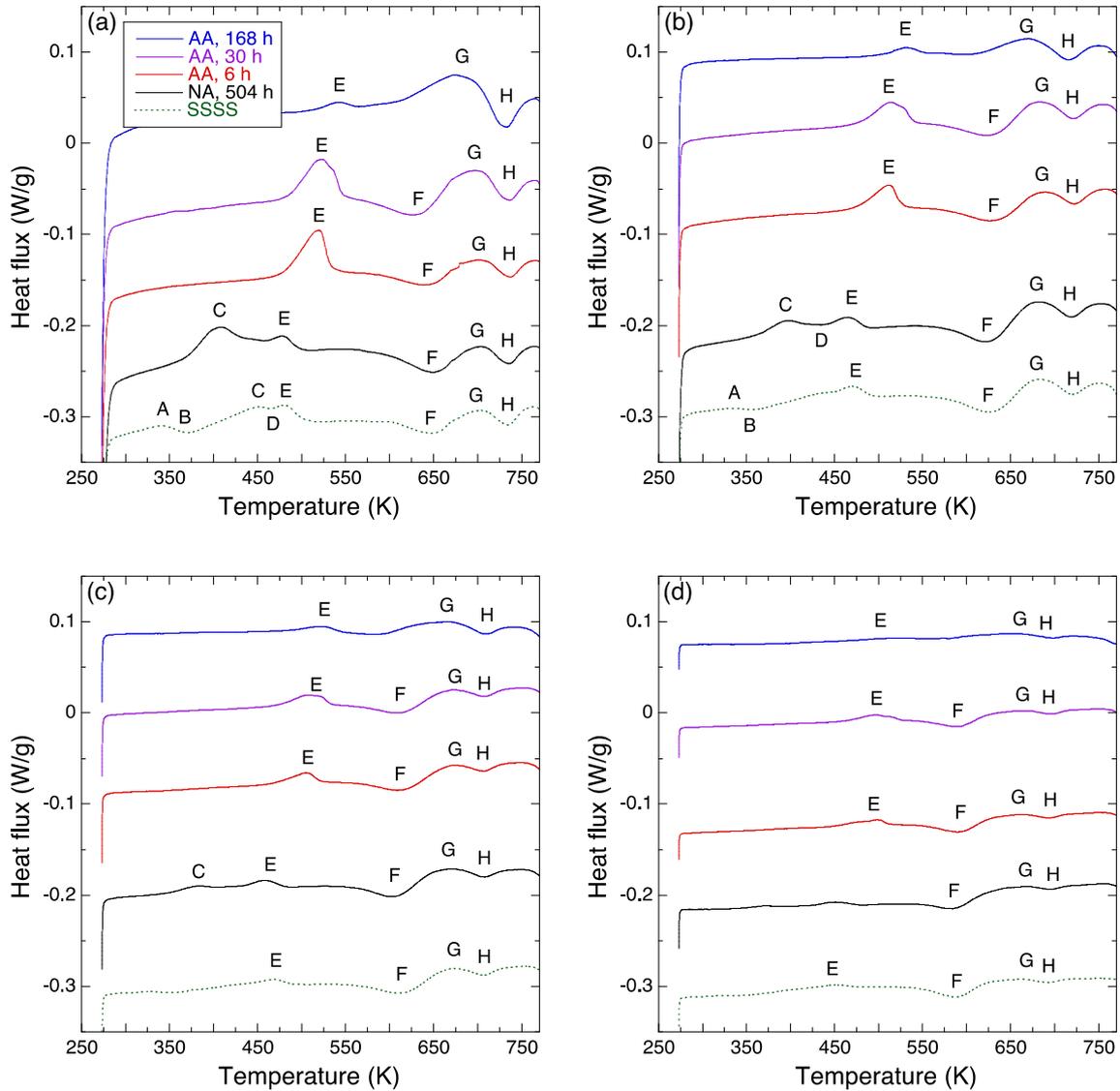

Figure 7 – Thermograms of the Al-4%Cu alloy after solution heat treatment (super saturated solid solution, SSSS), natural aging (NA) for 504 hours at 294K and artificial aging (AA) at 453 K for 6 h, 30 h and 168 h. (a) Heating rate of 20K/min. (b) Heating rate of 10K/min. (c) Heating rate of 5K/min. (d) Heating rate of 2.5K/min.

The $\theta''$ precipitates subsequently transformed into $\theta'$ particles (peaks E and F). The areas enclosed under peaks E and F reached a maximum at the peak-aged condition (at around 30 h) and then decreased progressively with aging time. These results are in excellent agreement with the TEM observations, which showed a maximum volume fraction of the $\theta''$ precipitates at 30 h and a progressive transformation of the $\theta''$ precipitates into $\theta'$ precipitates at longer times (Figs. 4 and 5). Finally, the area under the peaks G and H grew continuously with aging time as a larger volume fraction of $\theta'$ precipitates was present in the alloy and could be transformed into the $\theta$ precipitates.

The calorimetry curves were analysed to determine the temperature ranges of the reactions which define the precipitation sequence in the alloy. After that, the Kissinger peak method (eq. 5) was employed to estimate the activation energy of each phase transformation. Figure 8 illustrates an example of the Kissinger plot for the case of dissolution of the $\theta''$ precipitates in the solution treated sample (Peak E). Similar results were obtained for the other peaks from the thermograms obtained at different heating rates (from 2.5K/min to 20K/min) but they are not plotted for the sake of brevity. The activation energies of the different processes detected in the thermograms were obtained from the slope of the plot of $\ln(T_P^2/\beta)$ *vs* $1/RT_p$ and are summarized in Table 2 for the solution heat-treated and naturally aged specimens. They were quite consistent with the experimental results reported earlier by other authors [8, 12, 15, 20-23].

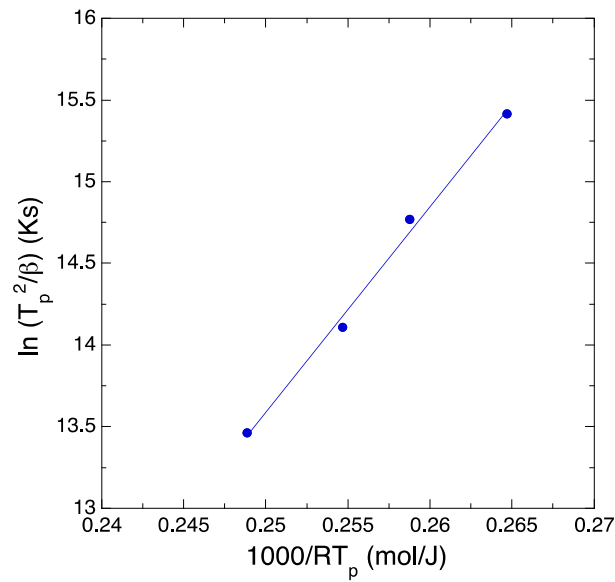

Figure 8 – Kissinger plot for the solution treated sample for the peak E (dissolution of the $\theta''$ precipitates). The straight line stands for the best fit of eq. (5) to the experimental data.

The temperature ranges of processes taking place in the solution treated alloy and the naturally aged alloy nearly coincide, with the exception of the dissolution of GP zones (Table 2). This process occured in the naturally aged specimen in the lower temperature range since GP zones were initially present in this condition. The activation energies of the processes taking place in the initial stages of the artificial aging treatment (including dissolution and precipitation of GP zones and $\theta''$ precipitates) were very close to the energy required for the movement of vacancies in the $\alpha$-Al matrix ($\sim 60$ kJ/mol) [24]. This result indicates that the migration of vacancies plays the key role in the development of the GP zones and of the $\theta''$ precipitates [25]. Further aging led to the $\theta'' \to \theta'$ phase transformation, a complex multistage reaction that involves more than one activation energy. The measured activation energies (related to peaks E and F) comprise information on three simultaneous processes: (a) the slow decomposition of the $\theta''$ phase, (b) the coarsening of the most stable $\theta''$ precipitates and (c) the heterogeneous precipitation of the $\theta'$ on dislocations and grain boundaries [13]. The activation energy associated with this complex reaction (around $112 - 126$ kJ/mol) is almost the same as the minimum energy required for the diffusion of Cu atoms in the Al matrix ($\sim 135$ kJ/mol) [8], suggesting that the whole process is driven by the migration of Cu atoms at equilibrium vacancy concentration. This alternative interpretation of the $\theta'' \rightarrow \theta'$ transition is fully consistent with the evolution of the size of thermal effect E (Fig. 6) and the activation energy related to the dissolution of the $\theta''$ plates (Table 2).

Table 2 – Temperature ranges and activation energies of the processes involved in the precipitation sequence of the Al-4%Cu alloy.

|  | Temperature range (K) | | Activation energy Q (kJ/mol) | |
| --- | --- | --- | --- | --- |
|  | Solution treated | Naturally aged 504 h at 293K | Solution treated | Naturally aged 504 h at 293K |
| Cu-clusters precipitation | – | – | – | – |
| Cu-clusters dissolution | 313 – 339 | – | 65 ± 5 | – |
| GP zones precipitation | 343 – 368 | – | 81 ± 6 | – |
| GP zones dissolution | 405 – 455 | 370 – 410 | 57 ± 7 | 58 ± 2 |
| $\theta''$ precipitation | 420 – 464 | – | 72 ± 5 | – |
| $\theta''$ dissolution | 454 – 483 | 451 – 479 | 126 ± 7 | 128 ± 5 |
| $\theta'$ precipitation | 588 – 641 | 584 – 646 | 112 ± 2 | 95 ± 2 |
| $\theta'$ dissolution | 664 – 706 | 648 – 705 | 172 ± 5 | 173 ± 6 |
| $\theta$ precipitation | 692 – 735 | 692 – 770 | 192 ± 1 | 195 ± 6 |
| $\theta$ dissolution | – | – | – | – |

The *in-situ* transformation $\theta'' \rightarrow \theta'$ was completed after 168 h of treatment, when no remaining $\theta''$ plates were found in the alloy (Fig. 4f). Longer aging times would result in the nucleation of the equilibrium $\theta$ phase. The transformation $\theta' \rightarrow \theta$ (peaks G and H) could solely be analysed by using the DSC thermograms, since no trace of the $\theta$ particles was found in the TEM observations because the aging temperature (453K) was too low to carry out for the $\theta' \rightarrow \theta$ transformation. It was observed that the energy barrier for such reaction (190-200 kJ/mol, Table 2) is somewhat lower than the values reported in the literature [12] (~ 300 kJ/mol) but substantially higher than the energy required for the diffusion of Cu atoms in the matrix. Two justifications for this fact are foreseen. First, purity of the examined material (99.9%) is greater compared to that of the alloy investigated in [12] (99.55%). It is well known that impurities (such as Fe) inhibit the mobility of Cu in the Al matrix at temperatures above ~ 300 °C, thus raising the energy involved in the migration of the alloying element [8]. Second, dislocations generated by the coarsening process of the $\theta'$ plates at these temperatures may also alter the energy barrier of the transformation [8].

### 4.2. Dilatometry analysis

The dilatation curves of solution treated specimens, expressed as the increment in length $\Delta L$ divided by the initial length $L_0$, are plotted in Fig. 9 during natural aging at 294K and artificial aging at 453K for ≈60 hours. The volume of the naturally aged specimen decreased slightly with aging time and remained constant after 50 hours. It is well known that alloying of Al with Cu atoms contracts the lattice parameters [9, 26]. Moreover, high resolution TEM analysis of the GP zones in [27] proved that the monolayer Cu atoms introduce displacements of the Al atoms around the GP zones leading to a further contraction of the alloy by the formation of Cu clusters and GP zones. The minor contraction observed in the dilation curve reflects this latter phenomenon.

The dilatometry curve of the material aged at 453K showed an initial reduction in volume, which can be attributed to the formation of the GP zones, in agreement with the TEM observations above. Afterwards, the specific volume of the alloy increased with aging time, in agreement with previous results in the literature [28]. This behaviour is mainly due to the reduction in the weight fraction of Cu in solid solution in the Al matrix as $\theta'$ and $\theta''$ precipitates are nucleated. The weight fraction of Cu in solid solution, $X_{Cu}$, was determined for each aging condition from the precipitate volume fraction $f$ in Table 1 and the atomic Cu content in each type of precipitate, which was 100% in GP zones, 25% in $\theta''$ ($Al_3Cu$) and 33% in $\theta'$ ($Al_2Cu$) [16]. The results are summarized in Table 3.

Table 3 – Weight percent of Cu dissolved in the matrix for different aging conditions

| Condition | $f$ (%) GP zones ($Cu$) | $f$ (%) $\theta''(Al_3Cu)$ | $f$ (%) of $\theta'(Al_2Cu)$ | $X_{Cu}$ (%) |
|---|---|---|---|---|
| SSSS | 0 | 0 | 0 | 4 |
| NA, 504 h at 294K | 1 | 0 | 0 | 3 |
| AA, 18 h at 453K | 0.4 | 0.3 | 0 | 3.5 |
| AA, 30 h at 453K | 0 | 1.6 | 0.6 | 2.9 |
| AA, 120 h at 453K | 0 | 1 | 0.8 | 3.1 |
| AA, 168 h at 453K | 0 | 0 | 1 | 3.5 |

The specific volume of the $\theta'$ ($Al_2Cu$) and $\theta''$ ($Al_3Cu$) can be found in Table 4. They were determined from the chemical composition and the lattice parameters for each crystal which were computed by means of Density Functional Theory (DFT). The details of the simulations as well as the actual values of the lattice parameters can be found in the Appendix. Moreover, the specific volume of the Al alloy decreased linearly with the weight % of Cu in solid solution. The proportionality constant was determined in [28] and can be found in Table 4. With this information as well as the data in Tables 3 and 4, it was possible to compute the expected volume increase in the Al-Cu alloy after 18 h of aging at 453K. The calculated increase in volume was in a good agreement with the dilatometry experiments and validates the information obtained by TEM on the volume fraction and size of the precipitates. It should be noted that the specific volumes of both $\theta''$ and the $\theta'$ precipitates are lower than that of the α-Al matrix, and the volume increase during precipitation was mainly controlled by the reduction in the weight fraction of Cu dissolved in the Al matrix as a result of precipitation [9].

Table 4 – Specific volumes of matrix and precipitates in the Al-4%Cu alloy.

| Phase | Specific volume ($cm^3/g$) |
|---|---|
| $\alpha$ | 0.3076 -7.35$X_{cu}$ |
| $\theta''$ | 0.2529 |
| $\theta'$ | 0.2447 |

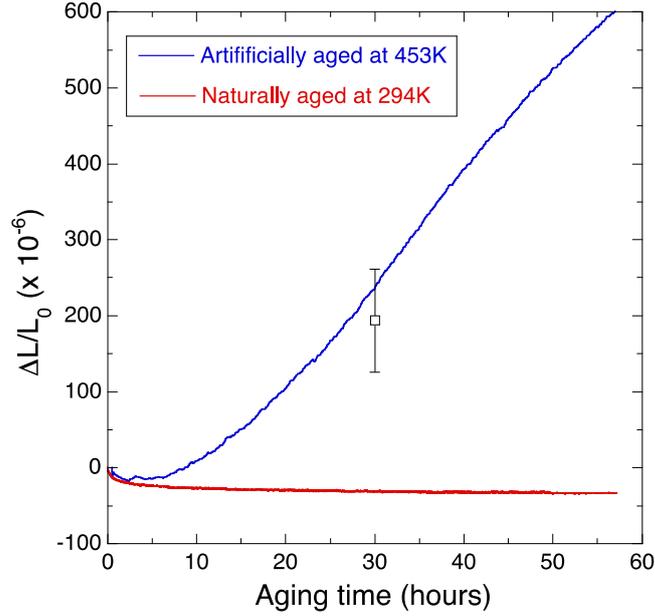

Figure 9 – Dilatometry curves of the solution treated Al-4%Cu alloy during natural aging at 294K and artificial aging at 453K. The square open symbol stands for the predictions obtained from the volume fractions of the different phases (α-Al matrix, $\theta''$, and $\theta'$) and their specific volumes taking into account the influence of the Cu atoms dissolved in the α-Al matrix.

### 4.3. Discussion

The information provided by the different characterization techniques provides a comprehensive picture of the precipitation process in this alloy. All the characterization techniques revealed the classical precipitation sequence at high temperature, supersaturated solid solution (SSSS) → GP zones → $\theta''$ → $\theta'$ → $\theta$. Nevertheless, the adjacent phase transformations were overlapped because of the complexity of the nucleation and precipitation phenomena, which depend on different parameters. GP zones were rapidly transformed at 453K into $\theta''$ precipitates, leading to a homogeneous distribution of the $\theta''$ precipitates in the matrix. The activation energies for the nucleation of GP zones and $\theta''$ precipitates were very close and similar to the energy required for the movement of vacancies in the α-Al matrix, which might be the rate limiting factor of the both processes. Moreover, the growth of the $\theta''$ phase was facilitated by the extremely low interfacial energy of the coherent $\theta''/\alpha-Al$ interface (6.6 mJ/m$^2$, see Appendix) although the $\theta'$ precipitate is more thermodynamically stable than the $\theta''$ (the formation enthalpy of $\theta'$ at 0 K was -15.41 kJ/mol while that of $\theta''$ was -8.3 kJ/mol, see Appendix).

The activation energy for the nucleation of the $\theta'$ phase was higher than that for the GP zones and $\theta''$ precipitates and similar to the minimum energy required for the diffusion of Cu atoms in the Al matrix. As a result, the $\theta'$ precipitates only appeared after long aging times at high temperature (30 h). Moreover, the nucleation of the $\theta'$ phase in the α-Al matrix is associated with a large stress-free transformation strain [13], which promoted their heterogeneous nucleation around dislocations and grain boundaries. Thus, the homogeneous nucleation of $\theta'$ precipitates (very likely from large $\theta''$ precipitates) only was detected after 120 h of aging. Finally, coarsening of the $\theta'$ precipitates was the domi-nant process between 120 h and 168 h of aging at 453K and was very likely responsible for the over-aging of the alloy.

## 5. Precipitate strengthening

The capability of the different precipitates to enhance the strength of the Al was evaluated by comparing the experimental microhardness with the predictions derived from theoretical models. The microhardness (HV) and critical resolved shear stress (CRSS) are shown in Table 5 for the different aging conditions. The conversion from microhardness to yield strength $\sigma_y$ was done by using the Tabor factor for plastic materials (T = 2.8) [29, 30], whereas the Critical Resolved Shear Stress (CRSS) (in MPa) was directly obtained from the yield strength by applying the Taylor factor for FCC polycrystals with random texture ($M = 3.06$) [3], leading to

$$\text{CRSS} = \frac{\sigma_y}{M} = \frac{T}{M}\text{HV} = 0.915 \text{ HV} \tag{6}$$

Table 5 – Microhardness and CRSS for the Al-4%Cu alloy in different aged conditions

| Condition | Microhardness (HV0.5) | CRSS (MPa) |
| --- | --- | --- |
| Solid Solution | $67 \pm 1.8$ | $61 \pm 1.6$ |
| Naturally aged at 293K | $96 \pm 1.5$ | $88 \pm 1.4$ |
| Artificially aged for 18 h at 453 K | $91 \pm 0.3$ | $83 \pm 0.3$ |
| Artificially aged for 30 h at 453 K | $108 \pm 2.0$ | $99 \pm 1.8$ |
| Artificially aged for 120 h at 453 K | $99 \pm 0.7$ | $91 \pm 0.6$ |
| Artificially aged for 168 h at 453 K | $87 \pm 2.4$ | $79 \pm 2.2$ |

According to the theoretical models [31, 32], there are four major contributions to the CRSS of the alloy at room temperature:

$$\text{CRSS} = \tau_0 + \tau_d + \tau_{ss} + \tau_p \tag{7}$$

where $\tau_0$ stands for the lattice resistance of pure Al, $\tau_d$ the strain hardening produced during indentation, $\tau_{ss}$ the solid solution strengthening and $\tau_p$ the precipitation hardening. Large plastic strains are attained under the indenter during the hardness test, and the contribution of strain hardening to the microhardness of the alloy cannot be neglected. However, this contribution – together with the lattice resistance $\tau_0$ – can be considered constant and independent of the alloy condition in a first approximation as the strain under the indenter is similar in all cases [29]. Under this assumption, $\tau_d + \tau_0 \approx \text{CRSS} - \tau_{ss}$ in the alloy in the supersaturated solid solution condition.

The solid solution contribution to the CRSS of the material, $\tau_{ss}$, was calculated based on the widely accepted relationship for FCC materials [33-36]:

$$\tau_{ss} = H X_{Cu}^n \tag{8}$$

where $X_{Cu}$ is the weight fraction of Cu dissolved in the Al matrix (Table 3) and $\tau_{ss}$ is expressed in MPa and $H$ and $n$ are two constants. Former studies performed by Zhu and Starink [35] on binary Al-Cu alloys with the different Cu content indicated that $n = 1$ and $H = 22$ MPa.

The strengthening provided by the precipitates depends both on the dislocation-precipitate orientation and on the nature of their interaction. The precipitates stand as obstacles to dislocations which move in the {111} <110> slip system of the FCC matrix (Fig. 10a). The $\theta'$ and $\theta''$ phases grow as plates in the {100} planes of the $\alpha$-Al matrix, thus presenting three different orientation variants (Fig. 10a). The matrix/precipitate interfaces parallel to the broad face of the disks (face 1 in Fig. 10c) are coherent while the perpendicular ones (face 2 in Fig. 10c) are semi-coherent. The three different orientation variants of the particles lead to two geometrical interactions with the dislocations gliding on the (111) slip plane along the [110] orientation (Fig. 10b). The Burgers vector $b$ of dislocation is parallel to the trace of the plate in the (111) plane in one orientation and forms an angle of 60º in the other two orientations.

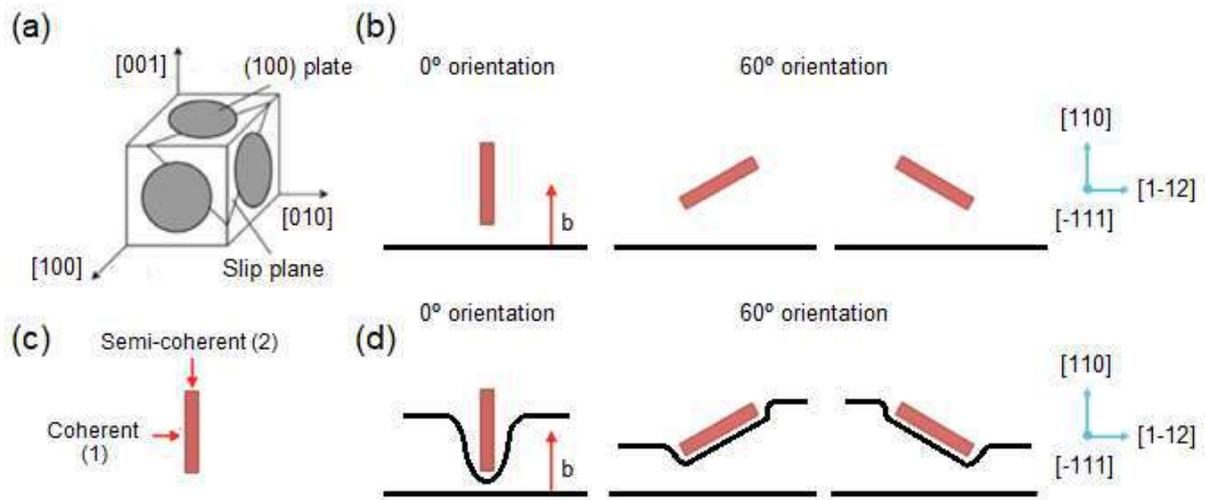

Fig. 10 – (a) Orientation variants of $\theta'$ and $\theta''$ precipitate plates in the FCC lattice of $\alpha$-Al. (b) Interfaces of the $\theta'$ and $\theta''$ precipitate plates. (c-d) Geometrical interaction of dislocations with precipitates in the (111) plane. The dislocation is represented by the black line.

Large $\theta'$ precipitates cannot be sheared by dislocations, which overcome the obstacle by the formation of an Orowan loop. Nie and Muddle [3] determined the CRSS necessary to overcome the platelet-shape precipitate by means of an Orowan loop, $\tau_{\theta'}$, taking into account the precipitate orientation with respect to the dislocation, which is given by,

$$\tau_{\theta'} = \frac{\mu b}{2\pi\sqrt{1-\nu}} \left( \frac{1}{0.931\sqrt{\frac{0.306\pi dt}{f}} - \frac{\pi d}{8} - 1.061t} \right) \ln\left(\frac{1.225t}{b}\right) \qquad (9)$$

where $\mu$ (= 26 GPa), $\nu$ (= 0.33) and $b$ (= 0.286 nm) stand for the shear modulus, the Poisson ratio and the Burgers vector of the α-Al matrix [3], $d$ is the diameter of the disk-shaped $\theta'$ precipitates, $t$ the thickness and $f$ the volume fraction. All the parameters for the $\theta'$ precipitates can be found in Table 1 for each aging condition.

Smaller $\theta''$ precipitates are assumed to be sheared by dislocations, and the shear stress to cut through the precipitate is given by [37]:

$$\tau_{\theta''} = \frac{0.908\, d}{t^2} \left( \frac{2\pi b f}{\mu b^2 \sqrt{\frac{d^2}{2b^2 f}}} \right)^{0.5} \gamma^{1.5} \tag{10}$$

where $\gamma$ stands for the interfacial energy between the matrix and the precipitate and the remaining parameters of the $\theta''$ precipitates can be found in Table 1 for the different aging conditions. Models to study dislocation shearing of precipitates assume a constant value of matrix/precipitate interface energy and do not take into account that interface energies depend on the particular interface orientation. This simplification may lead to significant differences in the strengthening provided by the precipitate depending on the dislocation/precipitate orientation if there is a large anisotropy in the interface energy. The interface energies corresponding to the coherent ($\gamma_c^{\theta''}$) and semi-coherent ($\gamma_{sc}^{\theta''}$) interfaces between the $\theta''$ precipitates and the α-Al matrix were determined by means of DFT in the Appendix (Table A3) and they show very large differences. Thus, an effective value of $\gamma$ was determined to be used in eq. (10) taking into account the geometrical interaction of the dislocations with the orientation variants.

One third of the precipitates are in the 0º orientation (Fig. 10) and the dislocation interacts with these precipitate variants through the semi-coherent interface 2. The corresponding interface energy is $\gamma_{sc}^{\theta''}$. Two thirds of the precipitates are in the 60º orientation (Fig. 10) and the dislocation will shear these precipitates through both the coherent interface 1 and semi-coherent interface 2. The averaged interface energy in this case is given by $\left[d^{\theta''}\gamma_c^{\theta''} + t^{\theta''}\gamma_{sc}^{\theta''}\right]/\left[d^{\theta''} + t^{\theta''}\right]$ where $d^{\theta''}$ and $t^{\theta''}$ stand for the diameter and the thickness of the $\theta''$ precipitates (Table 1).

Based on these geometrical considerations, an effective interface energy $\gamma_e^{\theta''}$ can be defined to compute the precipitate strengthening which is given by:

$$\gamma_e^{\theta''} = \frac{1}{3}\gamma_{sc}^{\theta''} + \frac{2}{3}\frac{\left[d^{\theta''}\gamma_c^{\theta''} + t^{\theta''}\gamma_{sc}^{\theta''}\right]}{\left[d^{\theta''} + t^{\theta''}\right]} \tag{11}$$

where the constants 1/3 and 2/3 stand for the fraction of precipitates oriented at 0° and 60°. This leads to an effective interface energy $\gamma_e^{\theta''}$ = 154 mJ/m², which can be introduced in eq. (10) to determine the value of $\tau_{\theta''}$.

The models presented above for $\theta'$ (non-shearable) and $\theta''$ (shearable) precipitates cannot be applied to the GP zones because, strictly speaking, such precipitates are not a phase. Although the mechanisms of the interaction of dislocations with GP zones have been studied using atomistic simulations [38], there is not a specific model to account for the actual strengthening provided by GP zones. The best available approximation considers the GP zones as spherical particles [31] of diameter $d$ and the CRSS to overcome the GP zone is given by

$$\tau_{GP} = \frac{1}{b}\sqrt{\frac{3f}{2\pi}}(0.72\mu b^2)\left(\frac{2}{d_c}\right)^{1.5}\left(\frac{d}{2}\right)^{0.5} \quad \text{if} \quad \frac{d}{2} < r_c \tag{12a}$$

$$\tau_{GP} = \frac{2}{db}\sqrt{\frac{3f}{2\pi}}(0.72\mu b^2) \quad \text{if} \quad \frac{d}{2} > r_c \tag{12b}$$

where $d_c$ = 20 nm is the critical precipitate diameter, which leads to a transition from dislocation shearing to bypassing [32]. Again, the volume fraction $f$ and the diameter $d$ of the GP zones for each aging condition can be found in Table 1.

Finally, the strengthening provided by a mixture of $n$ different types of precipitates, characterized by a critical resolved shear stress $\tau_i$ ($i = 1, ..., n$), can be determined as [39]:

$$\tau_p = \left\{\sum_{i=1}^{n}[\tau_i^q]\right\}^{1/q} \tag{13}$$

where the exponent 1≤ $q$ ≤ 2 depends on the relative strengths of the different types of obstacles. When all the obstacles have similar strength, as it is the case in the artificially aged Al-Cu alloy, $q$ = 2 [39].

The theoretical predictions of the CRSS obtained by adding the contributions of the lattice resistance, $\tau_0$, strain hardening, $\tau_d$, solid solution hardening, $\tau_{ss}$, and precipitate hardening, $\tau_p$, according to eq. (7), can be found in Table 6, together with the experimental results. The magnitudes of $\tau_{\theta''}$, $\tau_{\theta'}$ and $\tau_{GP}$ used to compute $\tau_p$ from eq. (13) are also included for comparison. As indicated above, $\tau_0 + \tau_d$ was assumed to be independent of the aging condition and was determined by subtracting the experimental value of the CRSS in the SSSS condition from both the lattice resistance and the solid solution hardening contributions.

A good agreement between the theoretical and experimental CRSS is observed (Table 6). For most of the conditions analyzed, the deviation between theoretical and experimental values is just about 3 %, and the maximum overestimation of the CRSS (by 9.3%) was found for the artificially aged alloy for 18 h. Therefore, it can be concluded that the applied models correctly predict the contributions of individual strengthening mechanisms. Analysis of these theoretical estimations provides interesting insights on the relative importance of the different contributions to the CRSS. Firstly, solid solution of the Cu atoms in the Al matrix ($\tau_{ss}$) is one of the main strengthening factors and does not change much during the heat treatments because the variations in weight fraction of Cu dissolved in the matrix in the NA and AA conditions is small (in the range 2.9%-3.5%, Table 5). Secondly, the largest contributions to the strength of the alloy are associated with a fine dispersion of small GP zones (in the naturally aged sample) or of $\theta''$(in the sample artificially aged for 30 h). The theoretical estimations of the strength provided by larger GP zones (in the sample artificially aged for 18 h) are very likely unrealistic because the model (eq. 12b) assumes that they cannot are bypassed by dislocations. However, good theoretical models for the strength provided by GP zones are not currently available. Thirdly, the best mechanical properties are obtained after 30 h of aging at 453K because of the synergistic contribution of the fact that presence of $\theta'$ and $\theta''$ precipitates. The theoretical model indicates that the contribution of the $\theta''$ precipitates is dominant with respect to that of the $\theta'$ and the difference is possible larger than that reported in Table 6 because the $\theta''$ precipitates were homogeneously distributed in the matrix while $\theta'$ plates tended to nucleate heterogeneously around dislocations and defects for his aging time and temperature. Further aging leads to a reduction in strength because the volume fraction of $\theta''$ precipitates decreases while the contribution of the $\theta'$ precipitates remains approximately constant up 120 h of aging and decreases afterwards because of the coarsening of the precipitate (Table 1). It should be finally noted that the strengthening provided by the homogeneous distribution of small GP zones obtained after naturally aging was comparable to the one provided by the $\theta''$ precipitates. Thus, small GP zones or $\theta''$ precipitates homogeneously distributed throughout the Al matrix seem to be the most efficient strengthening particles in contrast with larger $\theta'$ precipitates.

Table 6 – Strengthening contributions for different aging conditions. All values are in MPa.

| Condition | $\tau_0 + \tau_d$ | $\tau_{ss}$ | $\tau_p$ | $\tau_{GP}$ | $\tau_{\theta''}$ | $\tau_{\theta'}$ | Theoretical CRSS | Experimental CRSS |
|---|---|---|---|---|---|---|---|---|
| SSSS | 32.6 | 28.8 | - | - | - | - | 61.3 | 61.3 ± 1.6 |
| NA, 504 h | 32.6 | 21.6 | 36.9 | 36.9 | - | - | 91.1 | 88.1 ± 1.4 |
| AA, 18 h | 32.6 | 24.9 | 33.4 | 32.1 | 9.5 | - | 90.9 | 83.2 ± 0.3 |
| AA, 30 h | 32.6 | 21.4 | 42.1 | - | 35.4 | 22.7 | 96.1 | 98.8 ± 1.8 |
| AA, 120 h | 32.6 | 22.6 | 32.7 | - | 22.2 | 24.1 | 87.9 | 90.9 ± 0.6 |
| AA, 168 h | 32.6 | 25.0 | 16.8 | - | - | 16.8 | 74.3 | 79.4 ± 2.2 |

# 6. Conclusions

A multidisciplinary approach was applied to understand the precipitation kinetics and the effect of second phase precipitates on strength of an Al-4%Cu alloy including high resolution transmission electron microscopy, differential scanning calorimetry, high resolution dilatometry, density functional theory calculations, microhardness measurements and theoretical estimations of the strength for each type of precipitate. The combination of the different characterization and simulation strategies provided a coherent picture of the complex precipitation process as well as quantitative data of the structural and thermodynamic parameters. Furthermore, the roadmap presented in this investigation can be used for the design and optimization of complex alloys.

The analysis of the precipitation process during aging of the Al-4% Cu alloy showed the classical sequence, supersaturated solid solution (SSSS) → GP zones → $\theta''$ → $\theta'$ → $\theta$, although the adjacent phase transformations are overlapped because of the complexity of the nucleation and growth phenomena. The precipitation process at high temperature started by the nucleation and growth of GP zones and $\theta''$ precipitates, which was controlled by vacancy diffusion. $\theta''$ precipitates grew rapidly afterwards due to low energy of the coherent $\theta''/\alpha-Al$, leading to a homogeneous distribution of these precipitates after 30 h. Nucleation of the $\theta'$ was controlled by the diffusion of Cu atoms and initially occurred in dislocations and grain boundaries due to the stress free transformation strain associated with the nucleation of this phase. A homogeneous distribution of $\theta'$ precipitates was only achieved after long (120 h) aging times. Further aging led to the coarsening of the $\theta'$ precipitates in the overaged condition.

The different contributions to the strength of the Al alloys were ascertained using the current available models. In particular, an effective interface energy concept was introduced to estimate the contribution from the shearable $\theta''$ precipitates. This effective interface energy was estimated as a function of the interface energies (calculated using density functional theory) corresponding to the coherent and semi-coherent interfaces taking into account the geometrical interaction of the dislocations with the orientation variants.

It was found that the solid solution of the Cu atoms in the Al matrix was one of the main strengthening factors that did not change much with the heat treatments. The maximum precipitate strengthening was provided by the homogeneous distribution of $\theta''$ precipitates obtained in the peak-aged condition, followed by the homogeneous distribution of small GP zones obtained after natural aging. The impact of the $\theta'$ precipitates on the strength in the peak-aged condition was limited because they were mainly found at dislocations and grain boundaries. Further aging at high temperature beyond the peak-aged condition promoted the homogeneous nucleation of $\theta'$ precipitates but their contribution to the strength was hindered by the rapid coarsening of $\theta'$ precipitates.

## Acknowledgments

This investigation was supported by the European Research Council (ERC) under the European Union's Horizon 2020 research and innovation program (Advanced Grant VIRMETAL, grant agreement No. 669141). BB acknowledges the support from the Spanish Ministry of Education through the Fellowship FPU15/00403. IS acknowledges support by Madrid region under program S2013/MIT-2775, DIMMAT project. Useful discussions with Dr. Juan Pedro Fernandez are gratefully acknowledged.

**Appendix. Ab initio calculation of precipitate and interface properties**

$\theta''$ (Al$_3$Cu) and $\theta'$ (Al$_2$Cu) precipitates present a tetragonal structure. In the case of $\theta''$, the tetragonal lattice is formed by the substitution of a layer of Al atoms for Cu atoms in the (001) plane, leading to the unit cell depicted in Fig. A1(a). The body-center tetragonal structure of the $\theta'$ precipitates is found in Fig. A1(b).

The lattice parameters and elastic constants of $\theta''$ and $\theta'$ precipitates as well as the interfacial energies of the different interfaces between the precipitates and the α-Al matrix were determined using the density functional theory (DFT). The calculations were carried out with the help of the Quantum Espresso plane-wave pseudopotential code [40]. The exchange-correlation energy was evaluated employing the Perdew-Burke-Erzenhof (PBE) approach [41] within the Generalised Gradient Approximation (GGA). To reduce the basis set of plane wave functions used to describe the real electronic functions, ultrasoft pseudopotentials were used [42]. After careful convergence tests a cut off of 37 Ry (503 eV) was found to be sufficient to reduce the error in the total energy to less than 1 meV/atom. A k-point grid separation of 0.03 Å$^{-1}$ was employed for the integration over the Brillouin zone according to the Monkhorst–Pack scheme [43].

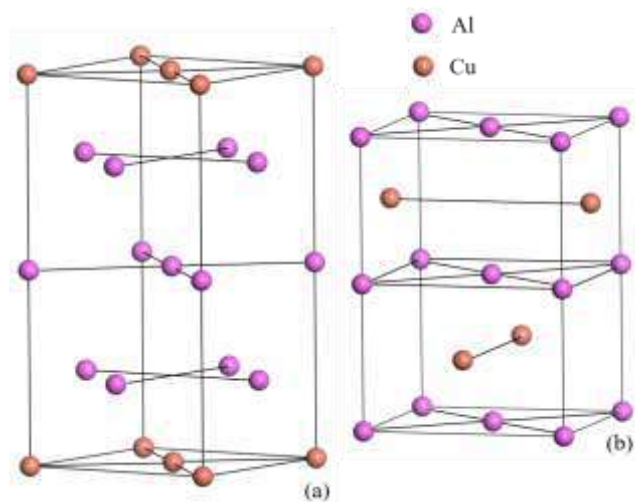

Fig. A1. (a) Unit cell of $\theta''$ (Al$_3$Cu) precipitate. (b) Unit cell of $\theta'$ (Al$_2$Cu) precipitate.

The lattice constants were obtained after relaxing the external and internal coordinates of the unit cell, thus reaching the ground state. The elastic constants were obtained by applying a given strain to the unit cell in the ground state and calculating the corresponding stress after the atom coordinates in the unit cell were relaxed. Taking into account the crystal symmetries, the cells of the $\theta''$ and $\theta'$ precipitates were deformed along two normal directions perpendicular to two faces of the tetragonal lattice and in two shear directions to compute the six independent elastic constants. Six strain levels (varying from -0.003 to 0.003) were used for each deformation pattern to obtain a reliable linear fit of the stress-strain relationship. The lattice and elastic constants obtained by DFT are shown in Tables A1 and A2, respectively. The lattice constants were very close to the experimental data in [44] and the values reported in [45]. Experimental values of the elastic constants are not available in the literature and the DFT values in Table A2 are reported here for the first time.

Table A1. Lattice parameters (in Å) of the $\theta'$ and $\theta''$ precipitates calculated by DFT.

| Phase | Lattice parameters | Experiments [44] | Reference [45] |
|---|---|---|---|
| $\theta'$ | a=b=4.05, c=5.82 | a=b=4.05, c=5.81 | - |
| $\theta''$ | a=b=3.98, c=7.66 | a=b=4.04, c=7.68 | a=b=3.9-3.975, c=7.525-7.666 |

Table A2. Elastic constants (in GPa) of the $\theta'$ and $\theta''$ precipitates calculated by DFT.

| Phase | $C_{11}=C_{22}$ | $C_{12}$ | $C_{13}=C_{23}$ | $C_{33}$ | $C_{44}=C_{55}$ | $C_{66}$ |
|---|---|---|---|---|---|---|
| $\theta'$ | 212.6 | 39.9 | 61.4 | 173.5 | 82.8 | 44.8 |
| $\theta''$ | 160.8 | 56.5 | 48.4 | 175.6 | 46.2 | 50.1 |

The interfacial energies were obtained from supercell calculations. N-atom supercells containing equal amounts of two phases A ($\alpha$-Al) and B (either $\theta'$ or $\theta''$) were considered. The $\theta'/\alpha$-Al and $\theta''/\alpha$-Al supercells are shown in Figs. A2 and A3, respectively. They were consistent with the experimental orientation relationship: $(001)_p\|\{001\}_{Al}$ and $[010]_p\|[010]_{Al}$, where p = $\theta'$ or $\theta''$ [46].

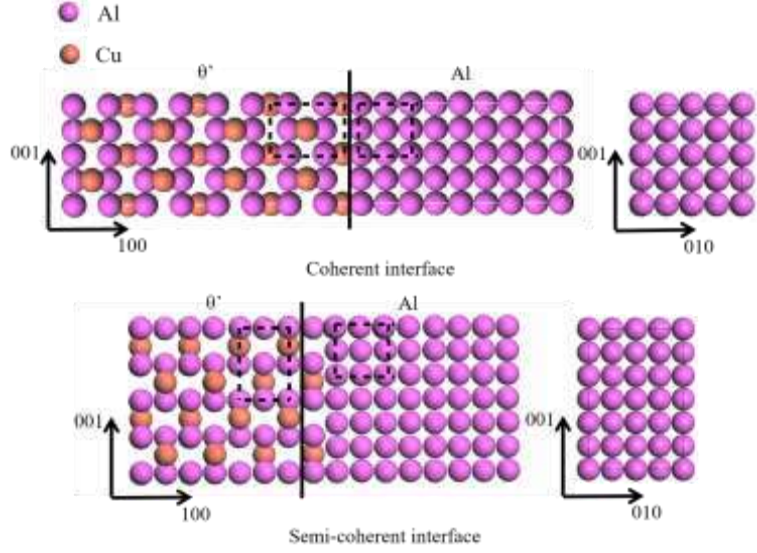

Figure A2. Configurations of $\theta'/\alpha$-Al coherent and semi-coherent interfaces. Dashed lines indicate the $1a_{\theta'} = 1a_{Al}$ and $2c_{\theta'} = 3a_\alpha$ relationships for the coherent and semi-coherent interfaces, respectively.

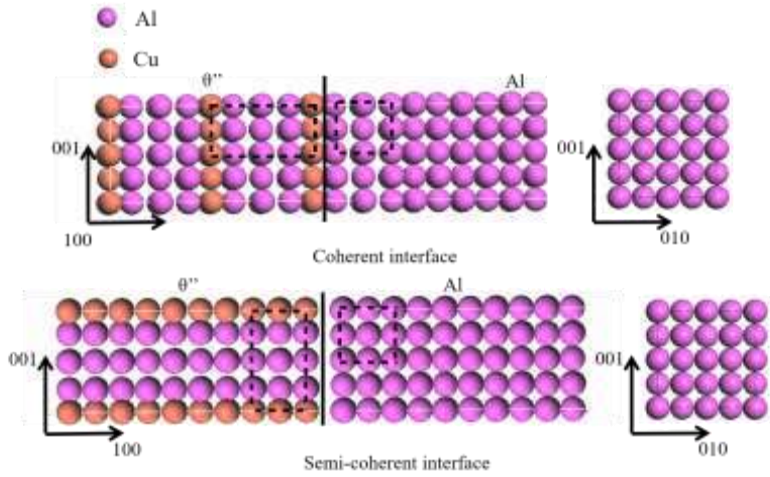

Figure A3. Configurations of $\theta''/\alpha$-Al coherent and semi-coherent interfaces. Dashed lines indicate the $1a_{\theta''} = 1a_{Al}$ and $1c_{\theta''} = 2a_\alpha$ relationships for the coherent and semi-coherent interfaces, respectively.

The formation energy of each $A_{N/2}B_{N/2}$ supercell with respect to the A and B phases, $\Delta E_{sc}$ (eV/atom) is given by

$$\Delta E_{sc} = E_{sc} - \frac{n}{N}E_A + \frac{m}{N}E_B \tag{A.1}$$

where $N$ is the total number of atoms in the supercell, $n$ and $m$ are the number of atoms of phases A and B, respectively, ($n + m = N$), and $E_A$ and $E_B$ stand for the formation energies of phases A and B (eV/atom), respectively, that are computed from unit cell calculations in the ground state. $\Delta E_{sc}$ is the extra energy that can be decomposed into two contributions: the coherency strain energy, $E_{cs}$, which is the strain energy required to accommodate the lattice mismatch between A and B and the interface energy, $\sigma$, which is the energy from atomic interactions at the interface [44]. $E_{cs}$ scales linearly with the volume of the supercell (for a fixed *n/m* ratio), while $\sigma$ is proportional to the interface area A. Thus,

$$\Delta E_{sc} = \frac{2\sigma A}{N} + E_{cs} \qquad (A.2)$$

and $\sigma$ can be obtained from the slope of the $\Delta E_{sc}$ *vs.* 1/N which is obtained from DFT simulations of the formation energy for supercells with different number of atoms.

The evolution of the formation energy, $\Delta E_{sc}$ *vs.* 1/N is plotted in Figure A4 for the four supercells with different interfaces and the interfacial energies for each interface were obtained from the slope. They are presented on Table A3. The results for the $\theta'$/*Al* interfaces are a very good agreement with the literature [46, 47] while those corresponding to both the coherent and semi-coherent interface between $\theta''$/*Al* at 0K are reported for the first time. While the interface energies of the semi-coherent interfaces are very close for both precipitates, a large difference was found between the energies of the $\theta'$/*Al* and $\theta''$/*Al* coherent interfaces. This is most possibly due to the fact that there is no significant difference between the structures of the matrix and the precipitate in the case of the $\theta''$/*Al* interface (it is essentially Cu (100) planes substituting Al), and no significant internal coordinate relaxation takes place during the formation of the interface. Furthermore, the small energy of the coherent $\theta'$/*Al* interface explains the formation of the $\theta''$ precipitate from the GP zones, although the $\theta'$ precipitate is more thermodynamically stable than the $\theta''$ (the formation enthalpy of $\theta'$ at 0K was -15.41 kJ/mol while that of $\theta''$ was -8.3 kJ/mol, according to the present study).

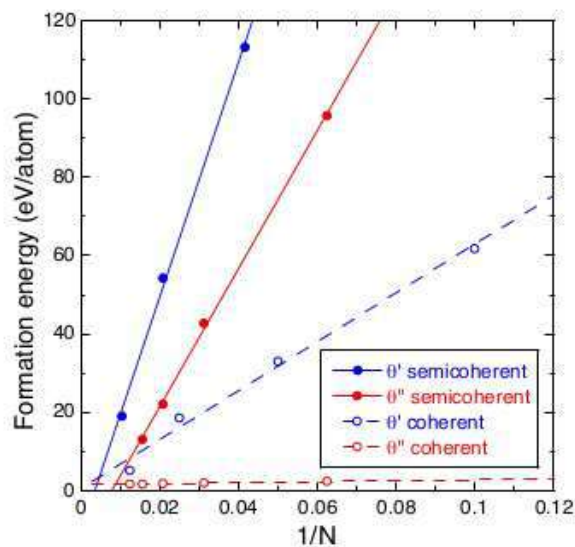

Figure A4. Formation energies of $\theta'$/*Al* and $\theta''$/*Al* supercells as a function of 1/N, where N is the number of atoms in the supercell for the coherent and semi-coherent configurations.

Table A3. Interfacial energies (in mJ/m$^2$) of $\theta''/Al$, $\theta'/Al$ coherent and semi-coherent interfaces obtained in this work and in references [46] and [47].

| Interface | Type | This work | [46] | [47] |
|---|---|---|---|---|
| $\theta'/Al$ | coherent | 152 | 235 | 170 |
| $\theta'/Al$ | semi-coherent | 487 | 615 | 520 |
| $\theta''/Al$ | coherent | 6.6 | - | - |
| $\theta''/Al$ | semi-coherent | 433 | - | - |